\documentclass[10pt,a4paper,onecolumn]{article}
\usepackage{marginnote}
\usepackage{graphicx}
\usepackage{xcolor}
\usepackage{authblk,etoolbox}
\usepackage{titlesec}
\usepackage{calc}
\usepackage{tikz}
\usepackage{hyperref}
\hypersetup{colorlinks,breaklinks=true,
            urlcolor=[rgb]{0.0, 0.5, 1.0},
            linkcolor=[rgb]{0.0, 0.5, 1.0}}
\usepackage{caption}
\usepackage{tcolorbox}
\usepackage{amssymb,amsmath}
\usepackage{ifxetex,ifluatex}
\usepackage{seqsplit}
\usepackage{xstring}

\usepackage{float}
\let\origfigure\figure
\let\endorigfigure\endfigure
\renewenvironment{figure}[1][2] {
    \expandafter\origfigure\expandafter[H]
} {
    \endorigfigure
}

\usepackage{fixltx2e} 
\usepackage[
  backend=biber,
]{biblatex}
\bibliography{paper.bib}

\let\textttOrig=\texttt
\def\texttt#1{\expandafter\textttOrig{\seqsplit{#1}}}
\renewcommand{\seqinsert}{\ifmmode
  \allowbreak
  \else\penalty6000\hspace{0pt plus 0.02em}\fi}

\makeatletter
\let\href@Orig=\href
\def\href@Urllike#1#2{\href@Orig{#1}{\begingroup
    \def\Url@String{#2}\Url@FormatString
    \endgroup}}
\def\href@Notdoi#1#2{\def\tempa{#1}\def\tempb{#2}%
  \ifx\tempa\tempb\relax\href@Urllike{#1}{#2}\else
  \href@Orig{#1}{#2}\fi}
\def\href#1#2{%
  \IfBeginWith{#1}{https://doi.org}%
  {\href@Urllike{#1}{#2}}{\href@Notdoi{#1}{#2}}}
\makeatother

\newlength{\cslhangindent}
\newlength{\csllabelwidth}
\newenvironment{CSLReferences}[3] 
 {
  \setlength{\parindent}{0pt}
  \ifodd #1 \everypar{\setlength{\hangindent}{\cslhangindent}}\ignorespaces\fi
  \ifnum #2 > 0
  \setlength{\parskip}{#2\baselineskip}
  \fi
 }%
 {}
\usepackage{calc}

\usepackage[top=3.5cm, bottom=3cm, right=1.5cm, left=1.0cm,
            headheight=2.2cm, reversemp, includemp, marginparwidth=4.5cm]{geometry}

\titleformat{\section}
  {\normalfont\sffamily\Large\bfseries}
  {}{0pt}{}
\titleformat{\subsection}
  {\normalfont\sffamily\large\bfseries}
  {}{0pt}{}
\titleformat{\subsubsection}
  {\normalfont\sffamily\bfseries}
  {}{0pt}{}
\titleformat*{\paragraph}
  {\sffamily\normalsize}

\usepackage{fancyhdr}
\makeatletter
\let\ps@plain\ps@fancy
\fancyheadoffset[L]{4.5cm}
\fancyfootoffset[L]{4.5cm}

\definecolor{linky}{rgb}{0.0, 0.5, 1.0}

\newtcolorbox{repobox}
   {colback=red, colframe=red!75!black,
     boxrule=0.5pt, arc=2pt, left=6pt, right=6pt, top=3pt, bottom=3pt}

\newcommand{\ExternalLink}{%
   \tikz[x=1.2ex, y=1.2ex, baseline=-0.05ex]{%
       \begin{scope}[x=1ex, y=1ex]
           \clip (-0.1,-0.1)
               --++ (-0, 1.2)
               --++ (0.6, 0)
               --++ (0, -0.6)
               --++ (0.6, 0)
               --++ (0, -1);
           \path[draw,
               line width = 0.5,
               rounded corners=0.5]
               (0,0) rectangle (1,1);
       \end{scope}
       \path[draw, line width = 0.5] (0.5, 0.5)
           -- (1, 1);
       \path[draw, line width = 0.5] (0.6, 1)
           -- (1, 1) -- (1, 0.6);
       }
   }

\patchcmd{\@maketitle}{center}{flushleft}{}{}
\patchcmd{\@maketitle}{center}{flushleft}{}{}
\patchcmd{\@maketitle}{\LARGE}{\LARGE\sffamily}{}{}
\def\maketitle{{%
  
  \AB@maketitle}}
\makeatletter
\renewcommand\AB@affilsepx{ \protect\Affilfont}
\renewcommand\AB@affilnote[1]{{\bfseries #1}\hspace{3pt}}
\renewcommand{\affil}[2][]%
   {\newaffiltrue\let\AB@blk@and\AB@pand
      \if\relax#1\relax\def\AB@note{\AB@thenote}\else\def\AB@note{#1}%
        \setcounter{Maxaffil}{0}\fi
        \begingroup
        \let\href=\href@Orig
        \let\texttt=\textttOrig
        \let\protect\@unexpandable@protect
        \def\thanks{\protect\thanks}\def\footnote{\protect\footnote}%
        \@temptokena=\expandafter{\AB@authors}%
        {\def\\{\protect\\\protect\Affilfont}\xdef\AB@temp{#2}}%
         \xdef\AB@authors{\the\@temptokena\AB@las\AB@au@str
         \protect\\[\affilsep]\protect\Affilfont\AB@temp}%
         \gdef\AB@las{}\gdef\AB@au@str{}%
        {\def\\{, \ignorespaces}\xdef\AB@temp{#2}}%
        \@temptokena=\expandafter{\AB@affillist}%
        \xdef\AB@affillist{\the\@temptokena \AB@affilsep
          \AB@affilnote{\AB@note}\protect\Affilfont\AB@temp}%
      \endgroup
       \let\AB@affilsep\AB@affilsepx
}
\makeatother

\renewcommand\Affilfont{\sffamily\small\mdseries}
\ifnum 0\ifxetex 1\fi\ifluatex 1\fi=0 
  \usepackage[T1]{fontenc}
  \usepackage[utf8]{inputenc}

\else 
  \ifxetex
    \usepackage{mathspec}
    \usepackage{fontspec}

  \else
    \usepackage{fontspec}
  \fi
  \defaultfontfeatures{Ligatures=TeX,Scale=MatchLowercase}

\fi
\IfFileExists{upquote.sty}{\usepackage{upquote}}{}
\IfFileExists{microtype.sty}{%
\usepackage{microtype}
\UseMicrotypeSet[protrusion]{basicmath} 
}{}

\usepackage{hyperref}
\hypersetup{unicode=true,
            pdftitle={XGA: A module for the large-scale scientific exploitation of archival X-ray astronomy data},
            pdfborder={0 0 0},
            breaklinks=true}
\let\addcontentslineOrig=\addcontentsline
\def\addcontentsline#1#2#3{\bgroup
  \let\texttt=\textttOrig\addcontentslineOrig{#1}{#2}{#3}\egroup}
\let\markbothOrig\markboth
\def\markboth#1#2{\bgroup
  \let\texttt=\textttOrig\markbothOrig{#1}{#2}\egroup}
\let\markrightOrig\markright
\def\markright#1{\bgroup
  \let\texttt=\textttOrig\markrightOrig{#1}\egroup}

\usepackage{graphicx,grffile}
\makeatletter
\def\maxwidth{\ifdim\Gin@nat@width>\linewidth\linewidth\else\Gin@nat@width\fi}
\def\maxheight{\ifdim\Gin@nat@height>\textheight\textheight\else\Gin@nat@height\fi}
\makeatother
\setkeys{Gin}{width=\maxwidth,height=\maxheight,keepaspectratio}
\IfFileExists{parskip.sty}{%
\usepackage{parskip}
}{
\setlength{\parindent}{0pt}
\setlength{\parskip}{6pt plus 2pt minus 1pt}
}
\ifx\paragraph\undefined\else
\let\oldparagraph\paragraph
\renewcommand{\paragraph}[1]{\oldparagraph{#1}\mbox{}}
\fi
\ifx\subparagraph\undefined\else
\let\oldsubparagraph\subparagraph
\renewcommand{\subparagraph}[1]{\oldsubparagraph{#1}\mbox{}}
\fi

\title{\texttt{XGA}: A module for the large-scale scientific exploitation of archival X-ray astronomy data}

        \author[1]{David J. Turner\footnote{david.turner@sussex.ac.uk}}
          \author[1]{Paul A. Giles}
          \author[1]{Kathy Romer}
          \author[1]{Violetta Korbina}
    
      \affil[1]{Department of Physics and Astronomy, University of
Sussex, Brighton, BN1 9QH, UK}
  \date{\vspace{-7ex}}

\begin{document}
\maketitle

\marginpar{

  \begin{flushleft}
  \sffamily\small

  {\bfseries DOI:} N/A

  \vspace{2mm}

  {\bfseries Software}
  \begin{itemize}
    \setlength\itemsep{0em}
    \item \href{https://github.com/DavidT3/XGA/}{\color{linky}{Repository}} \ExternalLink
    \item \href{https://xga.readthedocs.io/}{\color{linky}{Documentation}} \ExternalLink
  \end{itemize}

  \vspace{2mm}

  \par\noindent\hrulefill\par

  \vspace{2mm}


  {\bfseries Submitted:} N/A\\
  {\bfseries Published:} N/A

  \vspace{2mm}
  {\bfseries License}\\
  Authors of papers retain copyright and release the work under a Creative Commons Attribution 4.0 International License (\href{http://creativecommons.org/licenses/by/4.0/}{\color{linky}{CC BY 4.0}}).

  \end{flushleft}
}

\hypertarget{summary}{%
\section{Summary}\label{summary}}

The \emph{XMM} Cluster Survey (XCS, Romer et al., 2001) have developed a
new Python module, X-ray: Generate and Analyse (hereafter referred to as
\texttt{XGA}) to provide interactive and automated analyses of X-ray
emitting sources observed by the \emph{XMM}-Newton space telescope. \texttt{XGA} only requires that a set of cleaned, processed, event lists has been created, and (optionally) that a source detector has generated region lists for the observations.
\texttt{XGA} is centered around the concept of making all available data
easily accessible and analysable. The user provides information (e.g. RA, Dec, redshift) on the source they wish to investigate, and \texttt{XGA} will locate all
relevant observations and generate all required data products. This
approach means that the user can quickly and easily complete common
analyses without manually searching through large amounts of archival
data for relevant observations, thus being left free to focus on
extracting the maximum scientific gain. In the future, we will add support for X-ray telescopes other than {\em XMM} (e.g. {\em Chandra}, {\em eROSITA}), as well as the ability to perform multi-mission joint analyses. 
With the advent of new X-ray
observatories such as \emph{eROSITA} (Predehl et al., 2021),
\emph{XRISM} (XRISM Science Team, 2020), \emph{ATHENA} (Nandra et al.,
2013), and \emph{Lynx} (Gaskin et al., 2019), it is the perfect time for
a new, open-source, software package that is open for anyone to use and
scrutinise. 

\hypertarget{statement-of-need}{%
\section{Statement of need}\label{statement-of-need}}

X-ray telescopes allow for the investigation of some of the most extreme
objects and processes in the Universe; this includes galaxy clusters,
active galactic nuclei (AGN), and X-ray emitting stars. This makes the
analysis of X-ray observations useful for a variety of fields in
astrophysics and cosmology. Galaxy clusters, for instance, are useful as
astrophysical laboratories, and provide insight into how the Universe
has evolved during its lifetime.

Current generation X-ray telescopes have large archives of publicly
available observations; \emph{XMM}-Newton has been observing for over
two decades, for instance. This allows for analysis of
large amounts of archival data, but also introduces issues with respect
to accessing and analysing all the relevant data for a particular
source. \texttt{XGA} solves this problem by automatically identifying
the relevant \emph{XMM} observations then generating whatever data
products the user requires; from images to sets of annular spectra. Once
the user has supplied cleaned event lists (and optionally region files)
an analysis region can be specified and spectra (along with any
auxiliary files that are required) can be created.

Software to generate X-ray data products is supplied by the telescope
teams, and most commands require significant setup and configuration.
The complexity only increases when analysing multiple observations of a
single source, as is often the case due to the large archive of data
available. \texttt{XGA} provides the user with an easy way to generate
\emph{XMM} data products for large samples of objects (which will scale
across multiple cores), while taking into account complex factors (such
as removing interloper sources) that vary from source to source.

\begin{figure}
\centering
\includegraphics{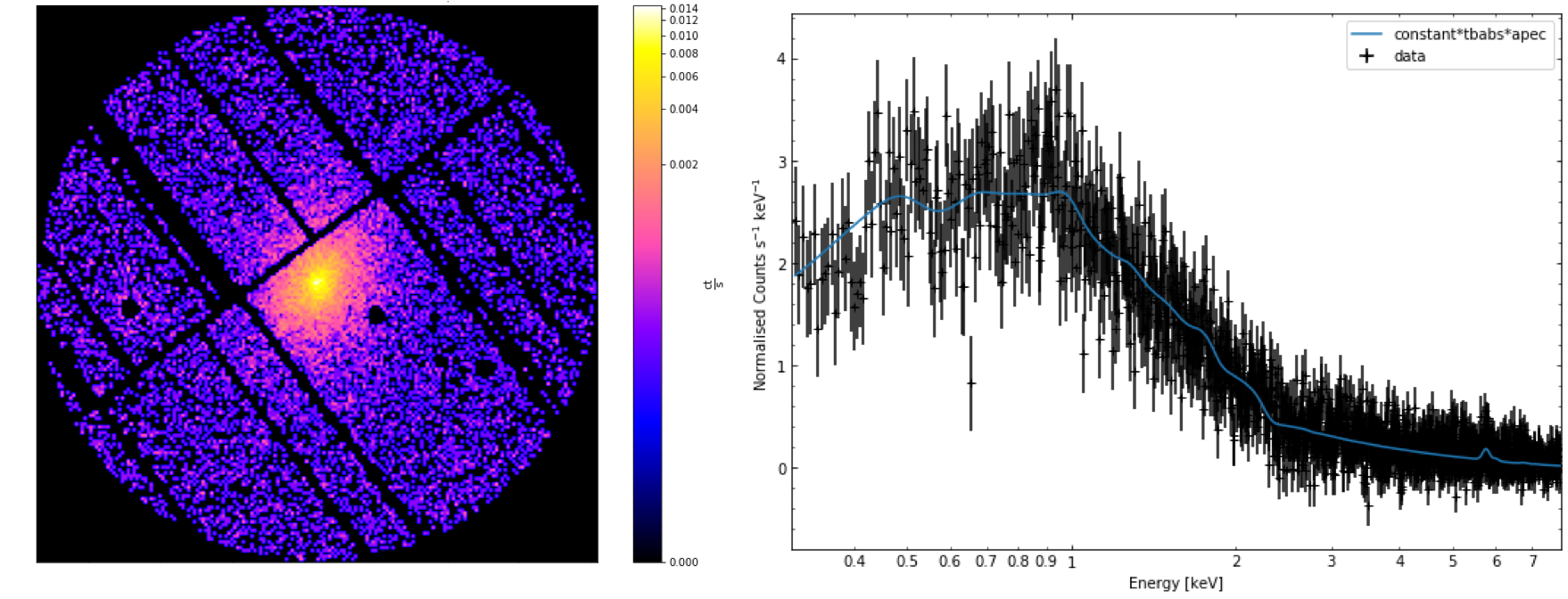}
\caption{Demonstration of the view methods of the \texttt{RateMap} and
\texttt{Spectrum} classes, when applied to the Abell 907 galaxy cluster.
Data from the \emph{XMM} EPIC-PN instrument of 0404910601 is used.
\emph{Left}: A count-rate map with a mask that removes contaminant
sources (using XCS region information) and applies an \(R_{500}\)
aperture. \emph{Right}: A spectrum generated for the \(R_{500}\) region
with contaminants removed, and fit with an absorbed plasma emission
model using XSPEC. \label{fig:rtspec}}
\end{figure}

\hypertarget{features}{%
\section{Features}\label{features}}

\texttt{XGA} is centered around \texttt{source} and \texttt{sample}
classes. Different \texttt{source} classes, which represent different
types of X-ray emitting astrophysical objects, all have different
properties and methods. Some properties and methods are common to all sources, but some store quantities or perform measurements that are only relevant to a particular type of astronomical source.

\texttt{XGA} also contains \texttt{product} classes, which provide
interfaces to X-ray data products, with built-in methods for analysis,
manipulation, and visualisation. The \texttt{RateMap} (a count rate map
of a particular observation) class for instance includes view methods
(left hand side of \autoref{fig:rtspec}), methods for coordinate
conversion, and for measuring the peak of the X-ray emission. We also
provide classes for interacting with, analysing, and viewing spectra
(see right hand side of \autoref{fig:rtspec}), both global and annular; as such we can use
\texttt{XGA} to investigate both average properties and, in the case of
extended sources, how these properties vary radially. Similar
procedures for image based analysis are also available, where images
(and merged images from all available data for a given source) can be
easily generated en masse, then combined with masks automatically
generated from supplied region files to perform photometric analyses.

We also include a set of profile classes, with built-in viewing methods,
and a fitting method based around the \texttt{emcee} ensemble MCMC
sampler (Foreman-Mackey et al., 2013). Profiles also support storing and
interacting with fitted models; including integration and
differentiation methods, inverse abel transforms, and predictions from
the model. An example of the utility of these profiles is the galaxy
cluster hydrostatic mass measurement feature; this requires the
measurement of 3D gas density profiles, 3D temperature profiles, gas
mass, and total mass profiles.

To extract useful information from the generated spectra, we implemented
a method for fitting models, creating an interface with XSPEC (Arnaud,
1996), a popular X-ray spectral fitting language. This interface
includes the ability to fit XSPEC models (e.g.~plasma emission and
blackbody) and simplifies interaction with the underlying software and
data by automatically performing simultaneous fits with all available
data. 

\begin{figure}
\centering
\includegraphics{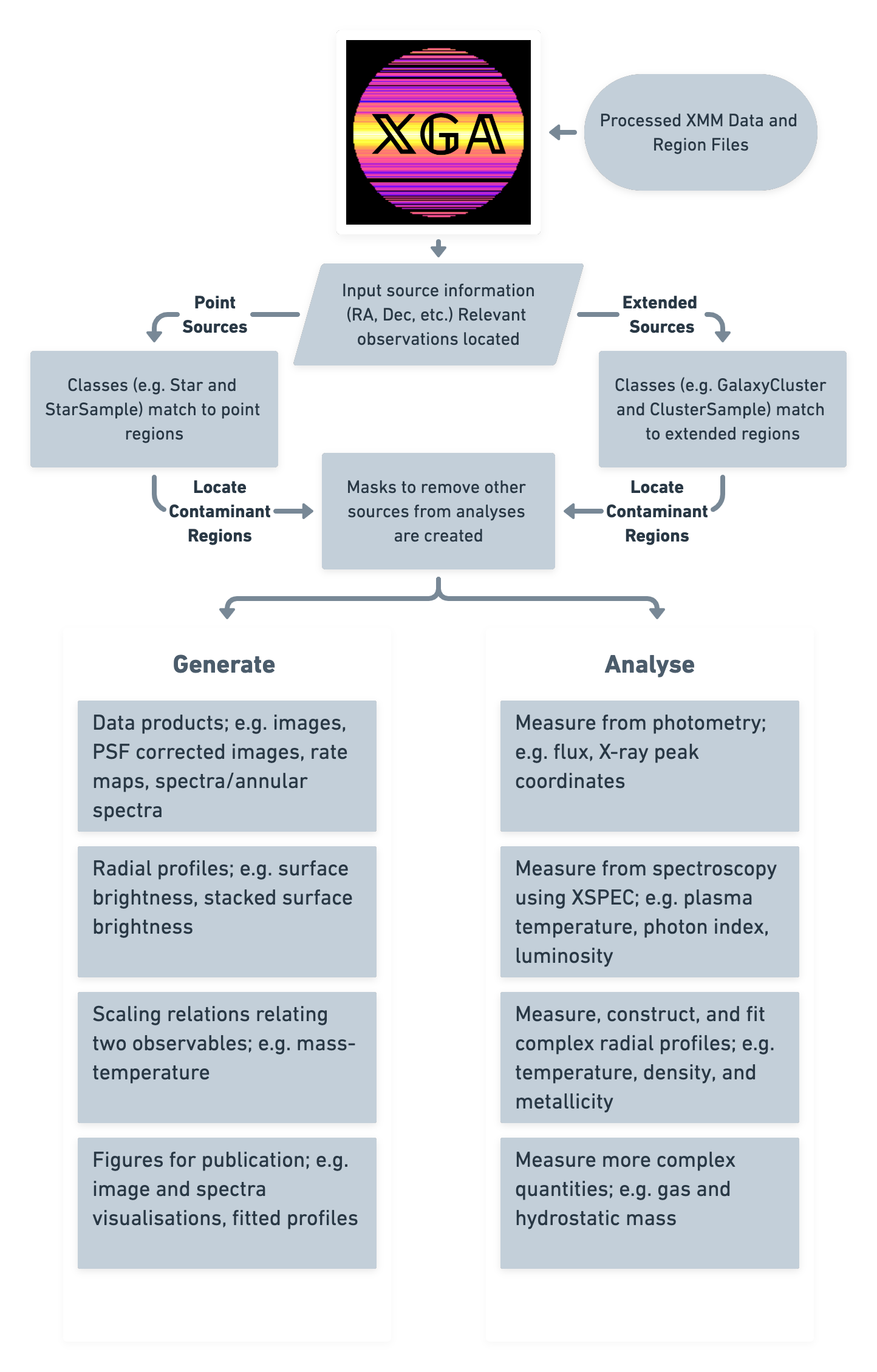}
\caption{A flowchart giving a brief overview of the \texttt{XGA}
workflow. \label{fig:flowchart}}
\end{figure}

\begin{figure}
\centering
\includegraphics{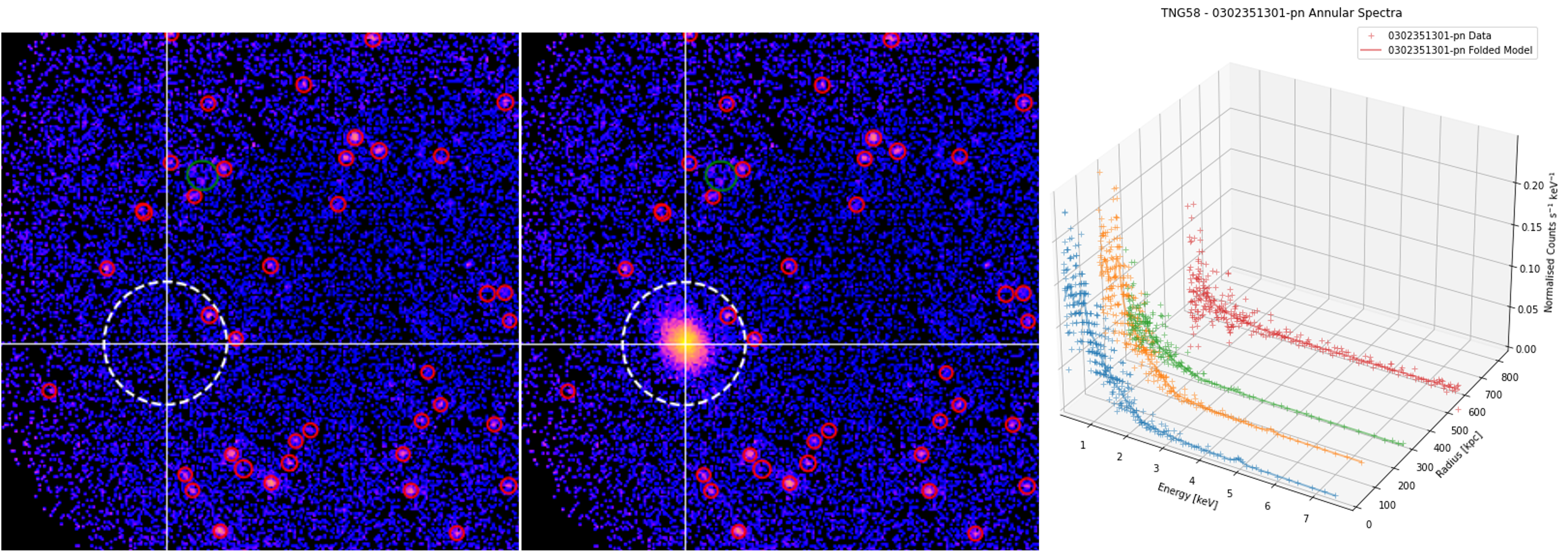}
\caption{A demonstration of the application of \texttt{XGA} to an artificial observation of a simulated Illustris-TNG cluster created by another XCS tool (XCSim; Turner et al. in prep). The simulated cluster was injected into the 0302351301 {\em XMM} observation. Cross-hair is the peak position of the simulated cluster in the middle image, dashed circle is the simulation-defined $R_{500}$ of the cluster. Red and green regions indicate XCS detected sources from the original observation. Left-hand-side: Zoomed view of the original combined PN+MOS1+MOS2 image. Middle: Zoomed view of the combined PN+MOS1+MOS2 image after the simulated cluster was added at $z=0.35$. Right-hand-side: An view of an \texttt{XGA} \texttt{AnnularSpectra} object generated from the simulated cluster. \label{fig:tng58}}
\end{figure}

\hypertarget{existing-software-packages}{%
\section{Existing software packages}\label{existing-software-packages}}

To the knowledge of the authors, no software package exists that
provides features completely equivalent to \texttt{XGA}, particularly in
the open source domain. That is not to say that there are no software
tools similar to the module that we have constructed; several research
groups including XCS (Lloyd-Davies et al., 2011), XXL (Giles et al.,
2016), LoCuSS (Martino et al., 2014), and the cluster group at UC Santa
Cruz (Hollowood et al., 2019) have developed pipelines to measure the
luminosity and temperature of X-ray emitting galaxy clusters, though
these have not been made public. It is also important to note that these
pipelines are normally designed to measure a particular aspect of a
particular type of X-ray source (galaxy clusters in these cases), and as
such they lack the generality and flexibility of \texttt{XGA}. Our new
software is also designed to be used interactively, as well as a basis
for building pipelines such as these.

Some specific analyses built into \texttt{XGA} have comparable open
source software packages available; for instance \texttt{pyproffit}
(Eckert et al., 2020) is a recently released Python module that was
designed for the measurement of gas density from X-ray surface
brightness profiles of galaxy clusters. We do not believe that any existing X-ray analysis
module has an equivalent to the source and sample based structure which
\texttt{XGA} is built around, or to the product classes that have been
written to interact with X-ray data products.

The \texttt{XSPEC} (Arnaud, 1996) interface we have developed for
\texttt{XGA} is far less comprehensive than the full Python wrapping
implemented in the \texttt{PyXspec} module, but scales with multiple
cores for the analysis of multiple sources simultaneously much more
easily.

\hypertarget{research-projects-using}{%
\section{\texorpdfstring{Research projects using
\texttt{XGA}}{Research projects using }}\label{research-projects-using}}

\texttt{XGA} is stable and appropriate for scientific use, and as such
it has been used in several recent pieces of work; this has included an
{\em XMM} analysis of the eFEDS cluster candidate catalogue (Turner et al.,
2021), where we produced the first temperature calibration between {\em XMM}
and {\em eROSITA}, a multi-wavelength analysis of an ACT selected galaxy
cluster (Pillay et al., 2021), and {\em XMM} follow-up of Dark Energy Survey
(DES) variability selected low-mass AGN candidates (Burke et al., 2021).

There are also several projects that use \texttt{XGA} nearing
publication. The first of these is a hydrostatic and gas mass analysis
of the redMaPPeR (Rykoff et al., 2014) SDSS selected XCS galaxy cluster
sample (Giles et al., 2022) and well as the ACTDR5 (Hilton et al.,
2021) Sunyaev-Zel'dovich (SZ) selected XCS sample of galaxy clusters.
This work also compares commonly measured X-ray properties of clusters
(the X-ray luminosity $L_{\rm{X}}$, and the temperature
$T_{\rm{X}}$ both to results from the existing XCS pipeline and from
literature, confirming that \texttt{XGA} measurements are consistent
with previous work. This process is repeated with \texttt{XGA}'s galaxy cluster gas and hydrostatic mass measurements, again showing they are consistent with previous work. \texttt{XGA}'s ability to stack and
combine X-ray surface brightness profiles is currently being used, in
combination with weak lensing information from DES, to look for signs of
modified gravity in galaxy clusters. Finally, we intend to use \texttt{XGA} to analyse artificial {\em XMM} observations of simulated galaxy clusters (see Figure~\ref{fig:tng58}) produced by another XCS tool (XCSim; Turner et al. in prep). This will allow us to perform measurements of simulated cluster properties for various simulation suites (e.g. Illustris-TNG, the 300 project) in an identical manner to our standard analyses. As such we shall be able to probe the realism of hydrodynamical simulations, as well as explore sources of measurement bias in X-ray observables.

\hypertarget{future-work}{%
\section{Future Work}\label{future-work}}

In the future we intend to introduce support for the analysis of X-ray
telescopes other than \emph{XMM}-Newton, first focusing on
\emph{Chandra} and \emph{eROSITA}, and then possibly considering the
addition of other X-ray instruments. This will include the same ability
to find relevant data and generate data products as is already
implemented for \emph{XMM}, and will also involve the introduction of
powerful multi-mission joint analyses to fully exploit the X-ray
archives.

We have already begun work that pairs \texttt{XGA} with XCS' simulated {\em XMM} observations (see Figure~\ref{fig:tng58}), and planned support for other telescopes will also extend to the analysis of simulated observations from future telescopes (e.g. {\em ATHENA} and {\em Lynx}). 

We are also happy to work with others to introduce specific analysis features that aren't already included in the module.

\hypertarget{acknowledgements}{%
\section{Acknowledgements}\label{acknowledgements}}

David J. Turner (DT), Kathy Romer (KR), and Paul A. Giles (PG)
acknowledge support from the UK Science and Technology Facilities
Council via grants ST/P006760/1 (DT), ST/P000525/1 and ST/T000473/1 (PG,
KR).

David J. Turner would like to thank Aswin P. Vijayan, Lucas Porth, and
Tim Lingard for useful discussions during the course of writing this
module.

We acknowledge contributions to the {\em XMM} Cluster Survey from A. Bermeo, M. Hilton, P. J. Rooney, S. Bhargava, L. Ebrahimpour, R. G. Mann, M. Manolopoulou, J. Mayers, E. W. Upsdell, C. Vergara, P. T. P. Viana, R. Wilkinson, C. A. Collins, R. C. Nichol, J. P. Stott, and others.

\hypertarget{references}{%
\section*{References}\label{references}}
\addcontentsline{toc}{section}{References}

\hypertarget{refs}{}
\begin{CSLReferences}{1}{0}
\leavevmode\hypertarget{ref-xspec}{}%
Arnaud, K. A. (1996). {XSPEC: The First Ten Years}. In G. H. Jacoby \&
J. Barnes (Eds.), \emph{Astronomical data analysis software and systems
v} (Vol. 101, p. 17).

\leavevmode\hypertarget{ref-desagn}{}%
Burke, C. J., Liu, X., Shen, Y., Phadke, K. A., Yang, Q., Hartley, W.
G., Harrison, I., Palmese, A., Guo, H., Zhang, K., Kron, R., Turner, D.
J., Giles, P. A., Lidman, C., Chen, Y.-C., Gruendl, R. A., Choi, A.,
Amon, A., Sheldon, E., \ldots{} DES Collaboration. (2021).
{Variability-Selected Dwarf AGNs in the Dark Energy Survey Deep Fields}.
\emph{arXiv e-Prints}, arXiv:2111.03079.
\url{http://arxiv.org/abs/2111.03079}

\leavevmode\hypertarget{ref-erositagasmass}{}%
Eckert, D., Finoguenov, A., Ghirardini, V., Grandis, S., Kaefer, F.,
Sanders, J. S., \& Ramos-Ceja, M. (2020). {Low-scatter galaxy cluster
mass proxies for the eROSITA all-sky survey}. \emph{arXiv e-Prints},
arXiv:2009.03944. \url{http://arxiv.org/abs/2009.03944}

\leavevmode\hypertarget{ref-emcee}{}%
Foreman-Mackey, D., Hogg, D. W., Lang, D., \& Goodman, J. (2013).
{emcee: The MCMC Hammer}. \emph{125}(925), 306.
\url{https://doi.org/10.1086/670067}

\leavevmode\hypertarget{ref-lynx}{}%
Gaskin, J. A., Swartz, D. A., Vikhlinin, A., Özel, F., Gelmis, K. E.,
Arenberg, J. W., Bandler, S. R., Bautz, M. W., Civitani, M. M.,
Dominguez, A., Eckart, M. E., Falcone, A. D., Figueroa-Feliciano, E.,
Freeman, M. D., Günther, H. M., Havey, K. A., Heilmann, R. K., Kilaru,
K., Kraft, R. P., \ldots{} Zuhone, J. A. (2019). {Lynx X-Ray
Observatory: an overview}. \emph{Journal of Astronomical Telescopes,
Instruments, and Systems}, \emph{5}, 021001.
\url{https://doi.org/10.1117/1.JATIS.5.2.021001}

\leavevmode\hypertarget{ref-xxllt}{}%
Giles, P. A., Maughan, B. J., Pacaud, F., Lieu, M., Clerc, N., Pierre,
M., Adami, C., Chiappetti, L., Démoclés, J., Ettori, S., Le Févre, J.
P., Ponman, T., Sadibekova, T., Smith, G. P., Willis, J. P., \& Ziparo,
F. (2016). {The XXL Survey. III. Luminosity-temperature relation of the
bright cluster sample}. \emph{592}, A3.
\url{https://doi.org/10.1051/0004-6361/201526886}

\leavevmode\hypertarget{ref-sdssxcs}{}%
Giles, P. A., Romer, A. K., Bermeo, A., \& Wilkinson, R. (2022).
\emph{{The XMM Cluster Survey analysis of the SDSS DR8 redMaPPer Catalogue: Implications for scatter, selection bias, and isotropy in cluster scaling relations.}} \emph{arXiv
e-Prints}, arXiv:2202.11107. \url{http://arxiv.org/abs/2202.11107}.

\leavevmode\hypertarget{ref-actdr5}{}%
Hilton, M., Sifón, C., Naess, S., Madhavacheril, M., Oguri, M., Rozo,
E., Rykoff, E., Abbott, T. M. C., Adhikari, S., Aguena, M., Aiola, S.,
Allam, S., Amodeo, S., Amon, A., Annis, J., Ansarinejad, B.,
Aros-Bunster, C., Austermann, J. E., Avila, S., \ldots{} Zhang, Y.
(2021). {The Atacama Cosmology Telescope: A Catalog of \textgreater4000
Sunyaev{{}}Zel{'}dovich Galaxy Clusters}. \emph{253}(1), 3.
\url{https://doi.org/10.3847/1538-4365/abd023}

\leavevmode\hypertarget{ref-matcha}{}%
Hollowood, D. L., Jeltema, T., Chen, X., Farahi, A., Evrard, A.,
Everett, S., Rozo, E., Rykoff, E., Bernstein, R., Bermeo-Hernandez, A.,
Eiger, L., Giles, P., Israel, H., Michel, R., Noorali, R., Romer, A. K.,
Rooney, P., \& Splettstoesser, M. (2019). {Chandra Follow-up of the SDSS
DR8 Redmapper Catalog Using the MATCha Pipeline}. \emph{244}(2), 22.
\url{https://doi.org/10.3847/1538-4365/ab3d27}

\leavevmode\hypertarget{ref-xcsmethod}{}%
Lloyd-Davies, E. J., Romer, A. K., Mehrtens, N., Hosmer, M., Davidson,
M., Sabirli, K., Mann, R. G., Hilton, M., Liddle, A. R., Viana, P. T.
P., Campbell, H. C., Collins, C. A., Dubois, E. N., Freeman, P.,
Harrison, C. D., Hoyle, B., Kay, S. T., Kuwertz, E., Miller, C. J.,
\ldots{} Stott, J. P. (2011). {The XMM Cluster Survey: X-ray analysis
methodology}. \emph{418}(1), 14--53.
\url{https://doi.org/10.1111/j.1365-2966.2011.19117.x}

\leavevmode\hypertarget{ref-locusshydro}{}%
Martino, R., Mazzotta, P., Bourdin, H., Smith, G. P., Bartalucci, I.,
Marrone, D. P., Finoguenov, A., \& Okabe, N. (2014). {LoCuSS:
hydrostatic mass measurements of the high-L\(_{X}\) cluster sample -
cross-calibration of Chandra and XMM-Newton}. \emph{443}(3), 2342--2360.
\url{https://doi.org/10.1093/mnras/stu1267}

\leavevmode\hypertarget{ref-athena}{}%
Nandra, K., Barret, D., Barcons, X., Fabian, A., den Herder, J.-W.,
Piro, L., Watson, M., Adami, C., Aird, J., Afonso, J. M., Alexander, D.,
Argiroffi, C., Amati, L., Arnaud, M., Atteia, J.-L., Audard, M.,
Badenes, C., Ballet, J., Ballo, L., \ldots{} Zhuravleva, I. (2013). {The
Hot and Energetic Universe: A White Paper presenting the science theme
motivating the Athena+ mission}. \emph{arXiv e-Prints}, arXiv:1306.2307.
\url{http://arxiv.org/abs/1306.2307}

\leavevmode\hypertarget{ref-denisha}{}%
Pillay, D. S., Turner, D. J., Hilton, M., Knowles, K., Kesebonye, K. C.,
Moodley, K., Mroczkowski, T., Oozeer, N., Pfrommer, C., Sikhosana, S.
P., \& Wollack, E. J. (2021). {A Multiwavelength Dynamical State
Analysis of ACT-CL J0019.6+0336}. \emph{Galaxies}, \emph{9}(4), 97.
\url{https://doi.org/10.3390/galaxies9040097}

\leavevmode\hypertarget{ref-erosita}{}%
Predehl, P., Andritschke, R., Arefiev, V., Babyshkin, V., Batanov, O.,
Becker, W., Böhringer, H., Bogomolov, A., Boller, T., Borm, K.,
Bornemann, W., Bräuninger, H., Brüggen, M., Brunner, H., Brusa, M.,
Bulbul, E., Buntov, M., Burwitz, V., Burkert, W., \ldots{} Yaroshenko,
V. (2021). {The eROSITA X-ray telescope on SRG}. \emph{647}, A1.
\url{https://doi.org/10.1051/0004-6361/202039313}

\leavevmode\hypertarget{ref-xcsfoundation}{}%
Romer, A. K., Viana, P. T. P., Liddle, A. R., \& Mann, R. G. (2001). {A
Serendipitous Galaxy Cluster Survey with XMM: Expected Catalog
Properties and Scientific Applications}. \emph{547}(2), 594--608.
\url{https://doi.org/10.1086/318382}

\leavevmode\hypertarget{ref-redmappersdss}{}%
Rykoff, E. S., Rozo, E., Busha, M. T., Cunha, C. E., Finoguenov, A.,
Evrard, A., Hao, J., Koester, B. P., Leauthaud, A., Nord, B., Pierre,
M., Reddick, R., Sadibekova, T., Sheldon, E. S., \& Wechsler, R. H.
(2014). {redMaPPer. I. Algorithm and SDSS DR8 Catalog}. \emph{785}, 104.
\url{https://doi.org/10.1088/0004-637X/785/2/104}

\leavevmode\hypertarget{ref-efedsxcs}{}%
Turner, D. J., Giles, P. A., Romer, A. K., Wilkinson, R., Upsdell, E.
W., Bhargava, S., Collins, C. A., Hilton, M., Mann, R. G., Sahl, M.,
Stott, J. P., \& Viana, P. T. P. (2021). {The XMM Cluster Survey: An
independent demonstration of the fidelity of the eFEDS galaxy cluster
data products and implications for future studies}. \emph{arXiv
e-Prints}, arXiv:2109.11807. \url{http://arxiv.org/abs/2109.11807}

\leavevmode\hypertarget{ref-xrism}{}%
XRISM Science Team. (2020). {Science with the X-ray Imaging and
Spectroscopy Mission (XRISM)}. \emph{arXiv e-Prints}, arXiv:2003.04962.
\url{http://arxiv.org/abs/2003.04962}

\end{CSLReferences}


@INPROCEEDINGS{xspec,
       author = {{Arnaud}, K.~A.},
        title = "{XSPEC: The First Ten Years}",
    booktitle = {Astronomical Data Analysis Software and Systems V},
         year = 1996,
       editor = {{Jacoby}, George H. and {Barnes}, Jeannette},
       series = {Astronomical Society of the Pacific Conference Series},
       volume = {101},
        month = jan,
        pages = {17},
       adsurl = {https://ui.adsabs.harvard.edu/abs/1996ASPC..101...17A}
}

@ARTICLE{xrism,
       author = {{XRISM Science Team}},
        title = "{Science with the X-ray Imaging and Spectroscopy Mission (XRISM)}",
      journal = {arXiv e-prints},
         year = 2020,
        month = mar,
          eid = {arXiv:2003.04962},
        pages = {arXiv:2003.04962},
archivePrefix = {arXiv},
       eprint = {2003.04962},
 primaryClass = {astro-ph.HE},
       adsurl = {https://ui.adsabs.harvard.edu/abs/2020arXiv200304962X}
}

@ARTICLE{athena,
       author = {{Nandra}, Kirpal and {Barret}, Didier and {Barcons}, Xavier and {Fabian}, Andy and {den Herder}, Jan-Willem and {Piro}, Luigi and {Watson}, Mike and {Adami}, Christophe and {Aird}, James and {Afonso}, Jose Manuel and {Alexander}, Dave and {Argiroffi}, Costanza and {Amati}, Lorenzo and {Arnaud}, Monique and {Atteia}, Jean-Luc and {Audard}, Marc and {Badenes}, Carles and {Ballet}, Jean and {Ballo}, Lucia and {Bamba}, Aya and {Bhardwaj}, Anil and {Stefano Battistelli}, Elia and {Becker}, Werner and {De Becker}, Micha{\"e}l and {Behar}, Ehud and {Bianchi}, Stefano and {Biffi}, Veronica and {B{\^\i}rzan}, Laura and {Bocchino}, Fabrizio and {Bogdanov}, Slavko and {Boirin}, Laurence and {Boller}, Thomas and {Borgani}, Stefano and {Borm}, Katharina and {Bouch{\'e}}, Nicolas and {Bourdin}, Herv{\'e} and {Bower}, Richard and {Braito}, Valentina and {Branchini}, Enzo and {Branduardi-Raymont}, Graziella and {Bregman}, Joel and {Brenneman}, Laura and {Brightman}, Murray and {Br{\"u}ggen}, Marcus and {Buchner}, Johannes and {Bulbul}, Esra and {Brusa}, Marcella and {Bursa}, Michal and {Caccianiga}, Alessandro and {Cackett}, Ed and {Campana}, Sergio and {Cappelluti}, Nico and {Cappi}, Massimo and {Carrera}, Francisco and {Ceballos}, Maite and {Christensen}, Finn and {Chu}, You-Hua and {Churazov}, Eugene and {Clerc}, Nicolas and {Corbel}, Stephane and {Corral}, Amalia and {Comastri}, Andrea and {Costantini}, Elisa and {Croston}, Judith and {Dadina}, Mauro and {D'Ai}, Antonino and {Decourchelle}, Anne and {Della Ceca}, Roberto and {Dennerl}, Konrad and {Dolag}, Klaus and {Done}, Chris and {Dovciak}, Michal and {Drake}, Jeremy and {Eckert}, Dominique and {Edge}, Alastair and {Ettori}, Stefano and {Ezoe}, Yuichiro and {Feigelson}, Eric and {Fender}, Rob and {Feruglio}, Chiara and {Finoguenov}, Alexis and {Fiore}, Fabrizio and {Galeazzi}, Massimiliano and {Gallagher}, Sarah and {Gandhi}, Poshak and {Gaspari}, Massimo and {Gastaldello}, Fabio and {Georgakakis}, Antonis and {Georgantopoulos}, Ioannis and {Gilfanov}, Marat and {Gitti}, Myriam and {Gladstone}, Randy and {Goosmann}, Rene and {Gosset}, Eric and {Grosso}, Nicolas and {Guedel}, Manuel and {Guerrero}, Martin and {Haberl}, Frank and {Hardcastle}, Martin and {Heinz}, Sebastian and {Alonso Herrero}, Almudena and {Herv{\'e}}, Anthony and {Holmstrom}, Mats and {Iwasawa}, Kazushi and {Jonker}, Peter and {Kaastra}, Jelle and {Kara}, Erin and {Karas}, Vladimir and {Kastner}, Joel and {King}, Andrew and {Kosenko}, Daria and {Koutroumpa}, Dimita and {Kraft}, Ralph and {Kreykenbohm}, Ingo and {Lallement}, Rosine and {Lanzuisi}, Giorgio and {Lee}, J. and {Lemoine-Goumard}, Marianne and {Lobban}, Andrew and {Lodato}, Giuseppe and {Lovisari}, Lorenzo and {Lotti}, Simone and {McCharthy}, Ian and {McNamara}, Brian and {Maggio}, Antonio and {Maiolino}, Roberto and {De Marco}, Barbara and {de Martino}, Domitilla and {Mateos}, Silvia and {Matt}, Giorgio and {Maughan}, Ben and {Mazzotta}, Pasquale and {Mendez}, Mariano and {Merloni}, Andrea and {Micela}, Giuseppina and {Miceli}, Marco and {Mignani}, Robert and {Miller}, Jon and {Miniutti}, Giovanni and {Molendi}, Silvano and {Montez}, Rodolfo and {Moretti}, Alberto and {Motch}, Christian and {Naz{\'e}}, Ya{\"e}l and {Nevalainen}, Jukka and {Nicastro}, Fabrizio and {Nulsen}, Paul and {Ohashi}, Takaya and {O'Brien}, Paul and {Osborne}, Julian and {Oskinova}, Lida and {Pacaud}, Florian and {Paerels}, Frederik and {Page}, Mat and {Papadakis}, Iossif and {Pareschi}, Giovanni and {Petre}, Robert and {Petrucci}, Pierre-Olivier and {Piconcelli}, Enrico and {Pillitteri}, Ignazio and {Pinto}, C. and {de Plaa}, Jelle and {Pointecouteau}, Etienne and {Ponman}, Trevor and {Ponti}, Gabriele and {Porquet}, Delphine and {Pounds}, Ken and {Pratt}, Gabriel and {Predehl}, Peter and {Proga}, Daniel and {Psaltis}, Dimitrios and {Rafferty}, David and {Ramos-Ceja}, Miriam and {Ranalli}, Piero and {Rasia}, Elena and {Rau}, Arne and {Rauw}, Gregor and {Rea}, Nanda and {Read}, Andy and {Reeves}, James and {Reiprich}, Thomas and {Renaud}, Matthieu and {Reynolds}, Chris and {Risaliti}, Guido and {Rodriguez}, Jerome and {Rodriguez Hidalgo}, Paola and {Roncarelli}, Mauro and {Rosario}, David and {Rossetti}, Mariachiara and {Rozanska}, Agata and {Rovilos}, Emmanouil and {Salvaterra}, Ruben and {Salvato}, Mara and {Di Salvo}, Tiziana and {Sanders}, Jeremy and {Sanz-Forcada}, Jorge and {Schawinski}, Kevin and {Schaye}, Joop and {Schwope}, Axel and {Sciortino}, Salvatore and {Severgnini}, Paola and {Shankar}, Francesco and {Sijacki}, Debora and {Sim}, Stuart and {Schmid}, Christian and {Smith}, Randall and {Steiner}, Andrew and {Stelzer}, Beate and {Stewart}, Gordon and {Strohmayer}, Tod and {Str{\"u}der}, Lothar and {Sun}, Ming and {Takei}, Yoh and {Tatischeff}, V. and {Tiengo}, Andreas and {Tombesi}, Francesco and {Trinchieri}, Ginevra and {Tsuru}, T.~G. and {Ud-Doula}, Asif and {Ursino}, Eugenio and {Valencic}, Lynne and {Vanzella}, Eros and {Vaughan}, Simon and {Vignali}, Cristian and {Vink}, Jacco and {Vito}, Fabio and {Volonteri}, Marta and {Wang}, Daniel and {Webb}, Natalie and {Willingale}, Richard and {Wilms}, Joern and {Wise}, Michael and {Worrall}, Diana and {Young}, Andrew and {Zampieri}, Luca and {In't Zand}, Jean and {Zane}, Silvia and {Zezas}, Andreas and {Zhang}, Yuying and {Zhuravleva}, Irina},
        title = "{The Hot and Energetic Universe: A White Paper presenting the science theme motivating the Athena+ mission}",
      journal = {arXiv e-prints},
         year = 2013,
        month = jun,
          eid = {arXiv:1306.2307},
        pages = {arXiv:1306.2307},
archivePrefix = {arXiv},
       eprint = {1306.2307},
 primaryClass = {astro-ph.HE},
       adsurl = {https://ui.adsabs.harvard.edu/abs/2013arXiv1306.2307N}
}

@ARTICLE{lynx,
       author = {{Gaskin}, Jessica A. and {Swartz}, Douglas A. and {Vikhlinin}, Alexey and {{\"O}zel}, Feryal and {Gelmis}, Karen E. and {Arenberg}, Jonathan W. and {Bandler}, Simon R. and {Bautz}, Mark W. and {Civitani}, Marta M. and {Dominguez}, Alexandra and {Eckart}, Megan E. and {Falcone}, Abraham D. and {Figueroa-Feliciano}, Enectali and {Freeman}, Mark D. and {G{\"u}nther}, Hans M. and {Havey}, Keith A. and {Heilmann}, Ralf K. and {Kilaru}, Kiranmayee and {Kraft}, Ralph P. and {McCarley}, Kevin S. and {McEntaffer}, Randall L. and {Pareschi}, Giovanni and {Purcell}, William and {Reid}, Paul B. and {Schattenburg}, Mark L. and {Schwartz}, Daniel A. and {Schwartz}, Eric D. and {Tananbaum}, Harvey D. and {Tremblay}, Grant R. and {Zhang}, William W. and {Zuhone}, John A.},
        title = "{Lynx X-Ray Observatory: an overview}",
      journal = {Journal of Astronomical Telescopes, Instruments, and Systems},
         year = 2019,
        month = apr,
       volume = {5},
          eid = {021001},
        pages = {021001},
          doi = {10.1117/1.JATIS.5.2.021001},
       adsurl = {https://ui.adsabs.harvard.edu/abs/2019JATIS...5b1001G}
}

@ARTICLE{erosita,
       author = {{Predehl}, P. and {Andritschke}, R. and {Arefiev}, V. and {Babyshkin}, V. and {Batanov}, O. and {Becker}, W. and {B{\"o}hringer}, H. and {Bogomolov}, A. and {Boller}, T. and {Borm}, K. and {Bornemann}, W. and {Br{\"a}uninger}, H. and {Br{\"u}ggen}, M. and {Brunner}, H. and {Brusa}, M. and {Bulbul}, E. and {Buntov}, M. and {Burwitz}, V. and {Burkert}, W. and {Clerc}, N. and {Churazov}, E. and {Coutinho}, D. and {Dauser}, T. and {Dennerl}, K. and {Doroshenko}, V. and {Eder}, J. and {Emberger}, V. and {Eraerds}, T. and {Finoguenov}, A. and {Freyberg}, M. and {Friedrich}, P. and {Friedrich}, S. and {F{\"u}rmetz}, M. and {Georgakakis}, A. and {Gilfanov}, M. and {Granato}, S. and {Grossberger}, C. and {Gueguen}, A. and {Gureev}, P. and {Haberl}, F. and {H{\"a}lker}, O. and {Hartner}, G. and {Hasinger}, G. and {Huber}, H. and {Ji}, L. and {Kienlin}, A. v. and {Kink}, W. and {Korotkov}, F. and {Kreykenbohm}, I. and {Lamer}, G. and {Lomakin}, I. and {Lapshov}, I. and {Liu}, T. and {Maitra}, C. and {Meidinger}, N. and {Menz}, B. and {Merloni}, A. and {Mernik}, T. and {Mican}, B. and {Mohr}, J. and {M{\"u}ller}, S. and {Nandra}, K. and {Nazarov}, V. and {Pacaud}, F. and {Pavlinsky}, M. and {Perinati}, E. and {Pfeffermann}, E. and {Pietschner}, D. and {Ramos-Ceja}, M.~E. and {Rau}, A. and {Reiffers}, J. and {Reiprich}, T.~H. and {Robrade}, J. and {Salvato}, M. and {Sanders}, J. and {Santangelo}, A. and {Sasaki}, M. and {Scheuerle}, H. and {Schmid}, C. and {Schmitt}, J. and {Schwope}, A. and {Shirshakov}, A. and {Steinmetz}, M. and {Stewart}, I. and {Str{\"u}der}, L. and {Sunyaev}, R. and {Tenzer}, C. and {Tiedemann}, L. and {Tr{\"u}mper}, J. and {Voron}, V. and {Weber}, P. and {Wilms}, J. and {Yaroshenko}, V.},
        title = "{The eROSITA X-ray telescope on SRG}",
      journal = {\aap},
         year = 2021,
        month = mar,
       volume = {647},
          eid = {A1},
        pages = {A1},
          doi = {10.1051/0004-6361/202039313},
archivePrefix = {arXiv},
       eprint = {2010.03477},
 primaryClass = {astro-ph.HE},
       adsurl = {https://ui.adsabs.harvard.edu/abs/2021A&A...647A...1P}
}

@ARTICLE{emcee,
       author = {{Foreman-Mackey}, Daniel and {Hogg}, David W. and {Lang}, Dustin and {Goodman}, Jonathan},
        title = "{emcee: The MCMC Hammer}",
      journal = {\pasp},
         year = 2013,
        month = mar,
       volume = {125},
       number = {925},
        pages = {306},
          doi = {10.1086/670067},
archivePrefix = {arXiv},
       eprint = {1202.3665},
 primaryClass = {astro-ph.IM},
       adsurl = {https://ui.adsabs.harvard.edu/abs/2013PASP..125..306F}
}
\end{document}